\let\ifarxiv=\iftrue
\pdfoutput=1

\ifarxiv
\documentclass[12pt,a4paper]{article}

\usepackage[nosort]{cite}
\usepackage{amsmath,amssymb}
\usepackage[a4paper,text={490pt,650pt},centering]{geometry}
\usepackage[explicit]{titlesec}

\titleformat{\section}[block]{\normalfont\bfseries\filcenter}{\scshape\thesection}{0.5em}{#1\\\titlerule}
\titlespacing{\section}{5pc}{*3}{*2}[5pc]

\makeatletter

\let\@authors\@empty
\let\@email\@empty
\let\@affiliation\@empty
\let\@pacs\@empty
\let\@keywords\@empty
\let\@preprint\@empty

\providecommand{\pacs}[1]{\gdef\@pacs{#1}}
\providecommand{\keywords}[1]{\gdef\@keywords{#1}}
\providecommand{\preprint}[1]{\toks@\expandafter{\@preprint#1\par}\edef\@preprint{\the\toks@}}
\renewcommand{\author}[1]{\ifx\@authors\@empty\toks@\expandafter{#1}\else\toks@\expandafter{\@authors, #1}\fi\edef\@authors{\the\toks@}}
\providecommand{\email}[1]{\ifx\@email\@empty\toks@\expandafter{#1}\else\toks@\expandafter{\@email, #1}\fi\edef\@email{\the\toks@}}
\providecommand{\affiliation}[1]{\gdef\@affiliation{#1}}

\makeatother
\fi

\ifarxiv\else

\PassOptionsToClass{aps,prl}{revtex4-1}
\iftrue
\PassOptionsToClass{nofootinbib,nobibnotes}{revtex4-1}
\PassOptionsToClass{showpacs,showkeys,preprintnumbers}{revtex4-1}
\fi
\documentclass[10pt,letterpaper,twocolumn,amsmath,amssymb,final]{revtex4-1}
\makeatletter
\let\o@a@f\@author@finish
\def\@author@finish{\o@a@f%
\let\@authors\empty\def\AF@opr##1{}\def\CO@opr##1##2##3{}%
\def\AU@opr##1##2##3{\ifx\@authors\@empty\toks@\expandafter{##2}\else\toks@\expandafter{\@authors, ##2}\fi\edef\@authors{\the\toks@}}
\@AAC@list}
\makeatother

\fi

\ifnum\pdfoutput=1\else
\PassOptionsToPackage{hypertex}{hyperref}
\PassOptionsToPackage{draft}{graphicx}
\fi

\usepackage[bookmarks=true,hyperfigures=true]{hyperref}
\usepackage{graphicx}

\allowdisplaybreaks[3]

\providecommand{\hypersetup}[1]{}

\hypersetup{plainpages=false}
\hypersetup{pdfpagemode=UseNone}
\hypersetup{bookmarksnumbered=true}
\hypersetup{pdfstartview=FitH}
\hypersetup{colorlinks=false}

\newcommand{\lrbrk}[1]{\left(#1\right)}

\def\<{\begin{eqnarray}}
\def\>{\end{eqnarray}}

\newcommand{\ddel}[2]{\frac{\partial #1}{\partial #2}}
\DeclareMathOperator{\tr}{tr}
\DeclareMathOperator{\Pf}{Pf}

\renewcommand{\[}{\begin{equation}}
\renewcommand{\]}{\end{equation}}

\newcommand{\C}{\mathbb{C}}
\newcommand{\nln}{\\\notag}

\begin{document}

\title{Subleading soft theorem in arbitrary dimension \ifarxiv\\\else\fi from scattering equations}

\author{Burkhard U.\ W.\ Schwab}
 \email{burkhard\_schwab@brown.edu}

\author{Anastasia Volovich}
 \email{anastasia\_volovich@brown.edu}

\affiliation{%
Brown University%
\ifarxiv\\\else\ (\fi%
Department of Physics%
\ifarxiv\\\else)\ \fi%
182 Hope St, Providence, RI, 02912%
}

% \begin{abstract}
% We investigate the new soft graviton theorem recently proposed in \href{http://www.arxiv.org/abs/1404.4091}{arXiv:1404.4091}.
% We use the CHY formula to prove this universal formula for both Yang-Mills theory and gravity scattering amplitudes at tree level in arbitrary dimension.
% \end{abstract}

\keywords{Scattering amplitudes, Yang-Mills theory, Gravity, soft theorem}
\pacs{04.50.+h,11.10.Kk,11.15.-q,11.55.Bq,11.80.Cr}

\date{\today}

%%%%%%%%%%%%%%%%%%%%%%%%%%%
% title page
\ifarxiv

\makeatletter
\thispagestyle{empty}
\begin{flushleft}
\footnotesize\ttfamily\@date
\end{flushleft}
\vspace{0.25cm}

\begin{centering}
\begingroup\Large\bf\@title\par\endgroup
\vspace{1cm}

\begingroup\@authors\par\endgroup
\vspace{5mm}

\begingroup\itshape\@affiliation\par\endgroup
\vspace{3mm}

\begingroup\ttfamily\@email\par\endgroup
\vspace{1cm}

\begin{minipage}{17cm}
 \begin{abstract}
 We investigate the new soft graviton theorem recently proposed in
 \href{http://arxiv.org/abs/1404.4091}{arXiv:1404.4091}.
 We use the CHY formula to prove this universal formula for both Yang-Mills theory and gravity scattering amplitudes at tree level in arbitrary dimension.
 \end{abstract}
%\theabstract
\end{minipage}
\vspace{1cm}

\end{centering}

\makeatother

%\tableofcontents

\fi

%%%%%%%%%%%%%%%%%%%%%%%%%%%%%%%%%%%%%%%%%%%%%%%%%%%%%%%%%%%%%%%%%%%%%%%%%%%%%%%%
% LETTER TITLE PAGE
\ifarxiv\else

\maketitle

\fi

%%%%%%%%%%%%%%%%%%%%%%%%%%%%%%%%%%%%%%%%%%%%%%%%%%%%%%%%%%%%%%%%%%%%%%%%%%%%%%%%
% COMMON BODY

\makeatletter
\hypersetup{pdftitle={\@title}}%
\hypersetup{pdfsubject={PACS numbers: \@pacs}}%
\hypersetup{pdfkeywords={\@keywords}}%
\hypersetup{pdfauthor={\@authors}}%
\makeatother

%%%%%%%%%%%%%%%%%%%%%%%%%%%%%%%%%%%%%%%%%%%%%%%%%%%%%%%%%%%%%%%%%%%%%%%%%%%%%%%%

\section{Introduction}
\label{sec:introduction}

Recently, Strominger proposed that a certain infinite-dimensional subgroup of the Bondi, van der Burg, Metzner, Sachs (BMS) supertranslation group is an exact symmetry of the quantum gravity $\mathcal{S}$-matrix \cite{Strominger:2013jfa}. Weinberg's soft theorem  \cite{Weinberg:1965nx,Weinberg:1964ew} is a Ward identity for this subgroup \cite{He:2014laa}. 

It was further conjectured by Cachazo and Strominger that there is a new soft graviton theorem \cite{Cachazo:2014fwa} which states that the subleading term $S^{(1)}_{\rm g}$ in the soft graviton expansion \[ M_{n+1}\to (S_{\rm g}^{(0)} + S_{\rm g}^{(1)} + S_{\rm g}^{(2)})M_{n}\label{gra}\] of the $(n+1)$-graviton scattering amplitude is also universal. Taking $q$ to be the momentum and $\epsilon_{\mu\nu}$ to be the polarization tensor of the soft particle, Weinberg's soft factor is \[S^{(0)}_{\rm g} = \sum_{a=1}^{n}\frac{\epsilon_{\mu\nu}k_{a}^{\mu}k_{a}^{\nu}}{q.k_{a}}\label{grb}\]  while the terms $S^{(1)}_{\rm g}$ and $S^{(2)}_{\rm g}$ are given by
\< S_{\rm g}^{(1)} = \sum_{a=1}^{n}\frac{\epsilon_{\mu\nu}k_{a}^{\mu}(q_{\lambda}J_{a}^{\lambda\nu})}{q.k_{a}},\ifarxiv\qquad\else\nln\fi S_{\rm g}^{(2)} = \sum_{a=1}^{n}\frac{\epsilon_{\mu\nu}(q_{\rho}J^{\rho\mu}_{a})(q_{\lambda}J_{a}^{\lambda\nu})}{q.k_{a}}.\label{eq:gravsub}\> Both subleading factors depend on the total angular momentum operator $J^{\mu\nu}_{a}=L_{a}^{\mu\nu} + S_{a}^{\mu\nu} =k_{a}^{[\mu}\ddel{}{k_{a,\nu]}} + ({\rm spin})$ of the $a$\textsuperscript{th} particle. In Yang-Mills theory a similar subleading factor was described in \cite{Casali:2014xpa} where the color-stripped amplitude has the soft limit
\[A_{n+1}\to (S_{\rm YM}^{(0)} + S_{\rm YM}^{(1)})A_{n},\] and the operators $S^{(0)}_{\rm YM}$ and $S^{(1)}_{\rm YM}$ are \[S^{(0)}_{\rm YM}=\sum_{\stackrel{\scriptstyle a\ {\rm adjacent\ } q}{\rm signed}}\frac{\epsilon.k_{a}}{q.k_{a}},\quad S^{(1)}_{\rm YM} = \sum_{\stackrel{\scriptstyle a\ {\rm adjacent\ } q}{\scriptsize\rm signed}}\frac{\epsilon_{\mu}q_{\nu}J_{a}^{\mu\nu}}{q.k_{a}}.\label{eq:ymsublead}\]
Gauge invariance of the Yang-Mills factor is derived from the antisymmetry in the indices. In gravity, gauge invariance of $S^{(1)}$ follows from global conservation of angular momentum while for $S^{(2)}$ it is the antisymmetry of $J^{\mu\nu}$. Relations
\eqref{gra}--\eqref{eq:gravsub}
 were proved in \cite{Cachazo:2014fwa,Casali:2014xpa} for tree level amplitudes using BCFW recursion relations \cite{Britto:2004ap,Britto:2005fq} in the spinor-helicity formalism  using a holomorphic limit as proposed in \cite{ArkaniHamed:2008gz}. Apart from BCFW, these technologies are not available in all dimensions. 

Cachazo, He, and Yuan (CHY) proposed a compact integral formula \cite{Cachazo:2013iea,Cachazo:2013hca} for tree-level scattering amplitudes of scalar $\phi^{3}$, (pure) Yang-Mills and gravity theories in arbitrary dimension. The amplitudes are given by an integral over points on a sphere which satisfy a set of algebraic equations, called the scattering equations. This formula generalizes the twistor string connected prescription for $\mathcal{N}=4$ SYM theory \cite{Roiban:2004yf} to scalar, gauge and gravity theories in arbitrary dimension.

In this \ifarxiv note \else letter \fi we perform a next-to-leading order expansion of the CHY integral in the presence of a soft particle in Yang-Mills theory and in gravity. This expansion can be used to compute the subleading soft factor for tree-level scattering amplitudes of these theories formulated on $d$-dimensional space-time. The subleading soft factors $S^{(1)}_{\rm g}$ and  $S^{(1)}_{\rm YM}$ take the same form in all dimensions. Given the momentum space form of both the subleading factors \eqref{eq:gravsub} and the CHY amplitude this is not unexpected because neither explicitly references dimension.  However, it is also surprising since the original conjecture of the universal subleading soft factors was based on the BMS symmetry principle which is only available in four dimensions. 

It would be very interesting to see how the subleading factors get corrected at loop level (see \cite{Dixon:2009ur,Dixon:2008gr,Bern:1998sv,Dunbar:2012aj} for loop corrections). Given the recent progress \cite{Geyer:2014fka,Mason:2013sva} in determining a stringy action principle for the derivation of the CHY form of scattering amplitudes, it might also be possible to determine the symmetry principle generating the subleading terms in dimensions other than four. We hope that further work will clarify these questions. 

\ifarxiv
This paper is organized as follows. In section \ref{sec:soft-fact-subl} we review the CHY tree-level formula for scattering amplitudes. Sections~\ref{sec:subl-soft-fact} and \ref{sec:subl-soft-fact-2} contain the calculation of the subleading terms in Yang-Mills theory and gravity in arbitrary dimension.
\fi

\section{Review of the CHY formula}
\label{sec:soft-fact-subl}

The CHY formula for tree-level scattering amplitudes \ifarxiv\else $M^{({\bf s})}_{n}$ \fi is
\[ \ifarxiv M^{(\bf s)}_{n} =\else\fi \int \frac{d^{n}\sigma}{{\rm vol} SL(2,\C)} {\prod_{a}}' \delta_{a}\left(\frac{\tr(T^{a_{1}}\cdots T^{a_{n}})}{\sigma_{12}\sigma_{23}\cdots\sigma_{n1}}\right)^{2-{\bf s}}\Pf'(\Psi)^{{\bf s}}\]
where the ``power'' ${\bf s}$ indicates whether the integral computes colored $\phi^{3}$ theory (${\bf s}=0$), (pure) Yang-Mills theory (${\bf s}=1$) or gravity (${\bf s}=2$) scattering amplitudes and $\sigma_{ij}=\sigma_{i}-\sigma_{j}$. In the following we frequently suppress the ${\rm vol}SL(2,\C)$. 
The $\delta$-distributions \[\delta_{a} = \delta\left(\sum_{b\neq a}\frac{k_{a}.k_{b}}{\sigma_{a}-\sigma_{b}}\right)\] impose $n-3$ \emph{scattering equations} \[\sum_{b\neq a}\frac{k_{a}.k_{b}}{\sigma_{a}-\sigma_{b}} = 0\] hence the primed product ${\prod_{a}}' = \sigma_{ij}\sigma_{jk}\sigma_{ki}\prod_{a\neq i,j,k}$ where $i,j,k$ may be chosen freely. The integral is to be taken over a sphere with $n$ punctures $\sigma_{i}$.  Due to $SL(2,\C)$-invariance, three of these $\sigma_{i}$ can be set to fixed values, s.t. the integral is $n-3$ dimensional. Therefore it is entirely fixed by the solutions to the scattering equations. We will be mainly interested in gravity (${\bf s}=2$) and gauge theory (${\bf s}=1$) as the soft limit of scalar amplitudes is identically zero already at leading order. In Yang-Mills theory, we will strip off the color factor $\tr(T^{a_{1}}\cdots T^{a_{n}})$ and work exclusively with color-ordered amplitudes to simplify the analysis. 

The factor $\Pf'\Psi$ is the Pfaffian of the $2n\times 2n$-dimensional matrix \[\Psi = \begin{pmatrix}A & -C^{T}\\ C & B\end{pmatrix}\] with \[A_{ab}=\begin{cases}\frac{k_{a}.k_{b}}{\sigma_{ab}} & a\neq b\\ 0 & a = b\end{cases},\quad C_{ab}=\begin{cases}\frac{\epsilon_{a}.k_{b}}{\sigma_{ab}} & a\neq b\\ -\sum_{c\neq a}\frac{\epsilon_{a}.k_{c}}{\sigma_{ac}} & a = b\end{cases}\] while $B$ looks just like $A$ with $k\to \epsilon$. Since the matrix $\Psi$ is actually singular, the Pfaffian is identically zero $\Pf\Psi=0$. The CHY integral prescribes therefore the use of the \emph{reduced Pfaffian} (indicated by the prime) \[\Pf'\Psi = 2\frac{(-1)^{i+j}}{\sigma_{ij}}\Pf\Psi^{ij}_{ij}\] where $\Psi^{ij}_{ij}$ is $\Psi$ with the $i$\textsuperscript{th} and $j$\textsuperscript{th} row and column removed. 

In \cite{Cachazo:2013hca} it was shown that the integral has the correct behavior in the limit of a graviton or a photon momentum going soft to leading order. We compute the next order factor following the steps already discussed in \cite{Cachazo:2013hca}. 

To begin, expand the product of $\delta$-distributions in the presence of a soft particle. We choose the $n$\textsuperscript{th} site to represent the soft particle, i.e. $k_n\to \varepsilon k_{n}$, $\varepsilon\ll1$. There are $(n-1)-3$ scattering equations which contain only one $k_{n}$ in the sum and one scattering equation which is proportional to $k_{n}$. As the variables in the $\delta$-distributions are complex and the integrand does not contain branch cuts, we may treat the $\delta$-distributions as poles. This allows us to rewrite the product ${\prod_{a}}'\delta_{a}$ as
\[\frac{1}{\varepsilon \sum_{b\neq n}\frac{k_{n}.k_{b}}{\sigma_{n}-\sigma_{b}}}{\prod_{a\neq n}}' \frac{1}{\varepsilon \frac{k_{a}.k_{n}}{\sigma_{a}-\sigma_{n}} + \sum_{b\neq a,n}\frac{k_{a}.k_{b}}{\sigma_{a}-\sigma_{b}}}.\] The second factor may be exactly expanded in a sum
\[\ifarxiv{\prod_{a\neq n}}' \frac{1}{\varepsilon \frac{k_{a}.k_{n}}{\sigma_{a}-\sigma_{n}} + \sum_{b\neq a,n}\frac{k_{a}.k_{b}}{\sigma_{a}-\sigma_{b}}} = \else\fi {\prod_{a\neq n}}'\sum_{i=0}^{\infty}\frac{\varepsilon^{i}}{i!}\left(\frac{k_{a}.k_{n}}{\sigma_{an}}\right)^{i}\delta^{(i)}\left(\sum_{b\neq a,n}\frac{k_{a}.k_{b}}{\sigma_{ab}}\right).\label{eq:deltaexp}\]
Here we denote $\delta^{(i)}(x) = \frac{(-1)^{i}i!}{x^{i+1}}$. For the Pfaffian factor, we may employ the useful expansion of the Pfaffian \[\Pf A = \sum_{\underset{q\neq p}{q=1}}^{2n} (-1)^{q} a_{p q} \Pf A^{pq}_{pq}.\label{eq:pfaffrecurse}\] The leading term in the expansion for $p=n$ is given by $q = 2n$ s.t. $\Pf' \Psi \to C_{nn} \Pf' \Psi^{n(2n)}_{n(2n)}$.  However, since we are also interested in the subleading terms, we need to look at the full expansion. As we expand along the row $p=n$ each coefficient apart from $q = 2n$ in the above expansion will be of order $\varepsilon$. More precisely, we have
\[\Pf' \Psi = -C_{nn}\Pf' \Psi^{n (2n)}_{n (2n)} + \varepsilon\sum_{\underset{q\neq i,j,n}{q=1}}^{2n-1} (-1)^{q}[\tilde{\Psi}^{ij}_{ij}]_{nq} \Pf' \Psi^{nq}_{nq}\label{eq:pfaffexp}\] with \[[\tilde{\Psi}^{ij}_{ij}]_{nq} = \begin{cases}\frac{k_{n}.k_{q}}{\sigma_{nq}} & q\leq n \\ \frac{k_{n}.\epsilon_{q-n}}{\sigma_{n(q-n)}} & q > n {\ \mbox{and}\ } q\neq 2n\end{cases}\] In \cite{Cachazo:2013iea} it was used that $\Pf' \Psi^{n(2n)}_{n(2n)}$ is actually independent of $k_{n}$ and $\epsilon_{n}$ and represents the correct factor for the $n-1$ particle amplitude \emph{to leading order}. At subleading order, there are still some terms proportional to $k_{n}$ in $\Pf \Psi^{n(2n)}_{n(2n)}$. The other Pfaffian minors similarly contain $k_{n}$ and $\epsilon_{n}$. Therefore, we will also have to expand these other minors along the row $(2n)$, i.e. we use \eqref{eq:pfaffrecurse} twice \[\Pf A = \ifarxiv \sum_{\underset{q\neq p}{q=1}}^{2n}\sum_{\underset{s\neq p,q,r}{s=1}}^{2n} (-1)^{q+s}a_{pq}a_{rs}\Pf (A^{pq}_{pq})^{rs}_{rs} =\else\fi \sum_{1\leq q < s \leq 2n}(-1)^{q+s}(a_{pq}a_{rs}-a_{ps}a_{rq})\Pf(A^{pq}_{pq})^{rs}_{rs}.\label{eq:unsav}\] This will \emph{not} provide us with another factor of $\varepsilon$. We will return to the importance of this second recursion and the subleading terms in $\Pf \Psi^{n(2n)}_{n(2n)}$ in due course. 

The interesting subleading soft factor in gauge theory and gravity is a single derivative operator \cite{Cachazo:2014fwa,Casali:2014xpa}. It can therefore be extracted from the expansion of the $\delta$-distributions and the leading terms in the Pfaffian recursion alone. However, the spin part $S^{\mu\nu}$ of the angular momentum operator $J^{\mu\nu}$ will not show up in this calculation since the scattering equations are explicitly independent of the polarization vectors $\epsilon_{a}$. The action of $S^{\mu\nu}$ can be seen in the Pfaffian expansion. 

The numerators of the soft factors follow from the leading order of the Pfaffian expansion \eqref{eq:pfaffexp}.  This was to be expected as the absence of the Pfaffian in the scalar case $({\bf s}=0)$ leads to the vanishing of the amplitude in the case of a soft particle. In the following section we investigate how the subleading soft factor emerges for gluon scattering amplitudes in arbitrary dimensions. Then, a similar analysis is done for graviton scattering amplitudes in the following section.

\section{Subleading soft factor in Yang-Mills theory}
\label{sec:subl-soft-fact}

The expansion in \eqref{eq:deltaexp} is exact. The $i=0$ case will lead us to the already known leading soft factor. For our purposes we need only the $i=1$ term in the expansion of the $\delta$-distribution. Since this provides us with a factor $\varepsilon$ we will not have to worry about subleading terms in the Pfaffian recursion for now. We write
\[\varepsilon\sum_{r\neq{i,j,k}}\left(\frac{k_{n}.k_{r}}{\sigma_{nr}}\right)\delta^{(1)}\left(\sum_{b\neq r,n}\frac{k_{r}.k_{b}}{\sigma_{rb}}\right){\prod_{a\neq{r,n}}}'\delta_{a}.\label{eq:deltafirst}\]
Notice firstly that the product now excludes the set $\{i,j,k,r,n\}$ of indices and secondly the presence of the derivative on one of the $\delta$s. Using only the leading factor coming from the Pfaffian ($C_{nn}\Pf' \Psi^{n(2n)}_{n(2n)}$) we may extract the subleading contribution \eqref{eq:ymsublead} to the soft factor for color-ordered gluon scattering amplitudes from
\[\int d^{n}\sigma\frac{\sum_{b\neq n}\frac{\epsilon_{n}.k_{b}}{\sigma_{nb}}}{\sum_{b\neq n}\frac{k_{n}.k_{b}}{\sigma_{nb}}}\frac{\sigma_{n-1,1}}{\sigma_{n-1,n}\sigma_{n,1}}{\sum_{r}}'\frac{k_{n}.k_{r}}{\sigma_{nr}}\delta^{(1)}_{r}{\prod_{a\neq r}}'\delta_{a}I_{n-1}\label{eq:softexp}.\] $I_{n-1}$ indicates that the rest of the integrand does not depend on $\sigma_{n}$. It is convenient to set $i=1$ and $j=n-1$ in the primed product and sum for the following calculation of the residues. We see that the integrand is regular for $\sigma_{n}\to \infty$ just as in the leading case. This allows us to treat the integral over $\sigma_{n}$ as in the leading case by deforming the contour. The only contributing poles are $\sigma_{n}=\sigma_{1}$, 
$\sigma_{n}=\sigma_{n-1}$ and also $\sigma_{n}=\sigma_{r}$ for every $r\neq 1,n-1,k$ ($k$ arbitrary) in the sum. In each case, the first fraction reduces to $\frac{\epsilon_{n}.k_{m}}{k_{n}.k_{m}}$ for $m = 1, n-1, r$. The second fraction is only interesting in the case $\sigma_{n}=\sigma_{r}$ as it is otherwise equal to unity. In the interesting case one rewrites \[\frac{\sigma_{n-1,1}}{\sigma_{n-1,r}\sigma_{r,1}} = \frac{\sigma_{n-1,r} + \sigma_{r1}}{\sigma_{n-1,r}\sigma_{r,1}}\] such that we get two terms. Putting this together, we find that \eqref{eq:softexp} reduces to a sum of four terms
\<\int d^{n-1}\sigma {\sum_{r\neq n}}'\Bigg\{\frac{\epsilon_{n}.k_{1}}{k_{n}.k_{1}}\frac{k_{n}.k_{r}}{\sigma_{1,r}}- \frac{\epsilon_{n}.k_{r}}{k_{n}.k_{r}}\frac{k_{n}.k_{r}}{\sigma_{1,r}} \ifarxiv\else\qquad{}\nln\fi+ \frac{\epsilon_{n}.k_{r}}{k_{n}.k_{r}}\frac{k_{n}.k_{r}}{\sigma_{n-1,r}} - \frac{\epsilon_{n}.k_{n-1}}{k_{n}.k_{n-1}}\frac{k_{n}.k_{r}}{\sigma_{n-1,r}}\Bigg\}\delta^{'}_{r}{\prod_{a\neq r}}'\delta_{a} I_{n-1}.\>
Now finally, we inspect the four terms and notice that they can be recovered from an operator of the form \[S^{(1)} = \frac{\epsilon_{n\mu}k_{n\nu} J^{\mu\nu}_{1}}{k_{n}.k_{1}} - \frac{\epsilon_{n\mu}k_{n\nu} J^{\mu\nu}_{n-1}}{k_{n}.k_{n-1}}\label{eq:subyang}\]
 acting on the product of $\delta$-distributions. This operator takes the exactly same form as the recently proposed operator for four dimensions \cite{Casali:2014xpa}.

As indicated above, finding the subleading soft factor from the Pfaffian recursion relation \eqref{eq:unsav} and the order $\varepsilon$ terms in the leading piece of the Pfaffian is more complicated. However, it can be done in the same way as above: First we use the leading order ($i=0$) in the expansion of the $\delta$-distributions. It is then necessary to write out all the order $\varepsilon$ terms in the Pfaffian. This contains the pieces from the leading order and also the pieces of order $\varepsilon$ from \eqref{eq:pfaffexp}. At this point it is useful to use \eqref{eq:unsav} with $p=n$ and $r=2n$ such that we get an expansion of the Pfaffian for $n-1$ particles. An investigation of the integrand will reveal once again that it is regular at infinity and there are no branch cuts, so a contour deformation is possible. After the integration the terms can be reassembled into the form of a sum of derivative operators $D$ acting on the Pfaffian $D \Pf A = \Pf A \tr(A^{-1}D A)$ where $A$ is the $n-1$ particle matrix $\Psi$. To see this, identities involving conservation of momentum contracted with polarization vectors $\epsilon_{i}$ and momenta $k_{i}$ have to be used. A final investigation will reveal that the terms can be recovered from the use of the operator \eqref{eq:subyang} on the $n-1$ particle Pfaffian. At this point, we can also see the action of the spin part $S^{\mu\nu}$ of the angular momentum operator, since there are additional terms in the Pfaffian expansion that cannot recovered from the action of the orbital angular momentum operator $L^{\mu\nu}$ alone.

\section{Subleading soft factor in gravity}
\label{sec:subl-soft-fact-2}

A very similar analysis can be done for graviton scattering amplitudes. We use the expansion of the $\delta$-distribution in \eqref{eq:deltafirst} and remember that since ${\bf s}=2$ we have two Pfaffian contributions. This implies that there is a pole for every particle in the amplitude
\[\int d^{n}\sigma \frac{\lrbrk{\sum_{b\neq n}\frac{\epsilon_{n}.k_{b}}{\sigma_{nb}}}^{2}}{\sum_{b\neq n}\frac{\epsilon_{n}.k_{b}}{\sigma_{nb}}}{\sum_{r}}'\frac{k_{n}.k_{r}}{\sigma_{nr}}\delta^{(1)}_{r}{\prod_{a\neq r}}'\delta_{a}I_{n-1}.\] However, due to the coefficient $\frac{k_{n}.k_{r}}{\sigma_{nr}}$ of the expansion of the $\delta$-distribution, there is no \emph{simple} pole at $\sigma_{n}=\sigma_{r}$ for every $r$. Instead, we find a double pole at this point. We will return to this issue shortly. Upon using the residue theorem in the above prescribed manner the simple poles yield a second sum $\sum_{b\neq r}$, s.t.
\[\int d^{n-1}\sigma {\sum_{r}}'\sum_{b\neq r} \frac{(\epsilon_{n}.k_{b})^{2}}{k_{n}.k_{b}}\frac{k_{n}.k_{r}}{\sigma_{br}}\delta_{r}^{(1)}{\prod_{a\neq r}}'\delta_{a} I_{n-1}.\label{eq:gravpart1}\] We may also consider the case of two Pfaffian factors which depend on different polarization vectors $\epsilon_{i}$ and $\tilde{\epsilon}_{i}$. Then $(\epsilon_{n}.k_{r})^{2}\to (\epsilon_{n}.k_{r})(\tilde{\epsilon}_{n}.k_{r})$ in the expressions above.

This is one part of the expression one would get from acting with an operator \[S^{(1)} = \sum_{a=1}^{n-1} \frac{\epsilon_{n}.k_{a}\epsilon_{n\mu}k_{n\lambda}J^{\lambda\mu}_{a}}{k_{n}.k_{a}}\label{eq:subgrav}\] on the product of $\delta$-distributions. Where is the rest? Quite interestingly, it hides in the second order poles. To find it we have to make use of Cauchy's integral formula
\[f^{(1)}(z_{0}) = \frac{1}{2\pi i}\oint \frac{f(z)dz}{(z-z_{0})^{2}}\] on the second order poles $\sigma_{n}=\sigma_{r}$ and act with a $\sigma_{n}$-derivative on the residue before setting $\sigma_{n}=\sigma_{r}$. After simplifying the result of the integration, one finds
\<\int d^{n-1}\sigma {\sum_{r}}'\sum_{b\neq r}\Bigg(\frac{(\epsilon_{n}.k_{r})^{2}k_{n}.k_{b}}{k_{n}.k_{r}\sigma_{rb}} - 2 \frac{\epsilon_{n}.k_{r}\epsilon_{n}.k_{b}}{\sigma_{rb}}\Bigg)\ifarxiv\else\nln\times\fi\delta_{r}^{(1)}{\prod_{a\neq r}}'\delta_{a} I_{n-1}\label{eq:gravpart2}\> Adding \eqref{eq:gravpart1} and \eqref{eq:gravpart2} produces all the terms expected from the action of operator \eqref{eq:subgrav} acting on the group of $\delta$-distributions. As we said before, we can also look at the result with two Pfaffians depending on different polarization vectors $\epsilon$ and $\tilde{\epsilon}$. This amounts to setting $(\epsilon_{n}k_{r})^{2}\to(\epsilon_{n}k_{r})(\tilde{\epsilon}_{n}k_{r})$ and $2(\epsilon_{n}k_{r})(\epsilon_{n}k_{b})\to (\epsilon_{n}k_{r})(\tilde{\epsilon}_{n}k_{b}) + \epsilon\leftrightarrow\tilde\epsilon$ in \eqref{eq:gravpart2}.

Notice that the primed product prevents us from seeing the subleading factor in the case of the four particle amplitude. This is in accord with the low-$n$ examples of the action of the subleading factor as given in \cite{Cachazo:2014fwa} and a nice check of the result. 

We performed the analysis for the Pfaffian-squared term in the integral. The major complication of the calculation for the Pfaffian factor w.r.t the Yang-Mills case is the presence of poles as well as double poles for every particle due to the interaction of the leading $C_{nn}\Pf'\Psi_{n-1}$ piece with the order $\varepsilon$ terms from the second Pfaffian present in the expression for gravity. However, the calculation uses the same technology as the Yang-Mills case.

\ifarxiv\paragraph{Acknowledgments.}\label{sec:acknowledgements}\else\begin{acknowledgments}\fi
We would like to thank Steven Avery, Miguel Paulos, Matteo Rosso, Marcus Spradlin, Congkao Wen and Michael Zlotnikov for useful discussions. We thank Freddy Cachazo for pointing out a missing term in the definition of the angular momentum operator in the previous version of the paper. This work is supported by the
US Department of Energy under contract DE-FG02-11ER41742 Early Career Award and the Sloan Research Foundation.
\ifarxiv\else\end{acknowledgments}\fi

%\ifarxiv\else\newpage\fi

\ifarxiv
\bibliographystyle{utphys}
\fi
\bibliography{BMSprj}

\end{document}